\documentclass[preprint,aps,epsfig]{revtex4}

\usepackage{psfig}

\def\lsim{\raise0.3ex\hbox{$<$\kern-0.75em\raise-1.1ex\hbox{$\sim$}}}
\def\gsim{\raise0.3ex\hbox{$>$\kern-0.75em\raise-1.1ex\hbox{$\sim$}}}

\newcommand{\be}{\begin{equation}}
\newcommand{\ee}{\end{equation}}
\def\alphaem{\alpha_{em}}

\def\beq{\begin{equation}}
\def\eeq{\end{equation}}
\def\beqa{\begin{eqnarray}}
\def\eeqa{\end{eqnarray}}

\newcommand{\rr}{\mbox{\boldmath $r$}}
\newcommand{\rrn}{\mbox{$r$}}

\newcommand{\rb}{\mbox{\boldmath $b$}}

\def\gappeq{\mathrel{\rlap {\raise.5ex\hbox{$>$}}
{\lower.5ex\hbox{$\sim$}}}}

\def\lappeq{\mathrel{\rlap{\raise.5ex\hbox{$<$}}
{\lower.5ex\hbox{$\sim$}}}}

\def\Toprel#1\over#2{\mathrel{\mathop{#2}\limits^{#1}}}

\begin{document}

\title{Probing the  Color Glass Condensate in an electron-ion collider} 
\author{  M.S.Kugeratski$^1$, V.P. Gon\c{c}alves$^2$, and  F.S. Navarra$^1$}
\affiliation{$^1$Instituto de F\'{\i}sica, Universidade de S\~{a}o Paulo, 
C.P. 66318,  05315-970 S\~{a}o Paulo, SP, Brazil\\
$^2$High and Medium Energy Group (GAME), \\
Instituto de F\'{\i}sica e Matem\'atica,  Universidade
Federal de Pelotas\\
Caixa Postal 354, CEP 96010-900, Pelotas, RS, Brazil\\}
\begin{abstract}

Perturbative Quantum Chromodynamics (pQCD) predicts that the small-$x$ gluons in a 
hadron wavefunction should form a Color Glass Condensate (CGC), characterized by a saturation scale $Q_s (x, A)$ which is energy and atomic number dependent. In this paper we study the predictions of CGC physics for  electron - ion collisions at high energies. We consider that the nucleus at high energies acts as an amplifier of the physics of high parton densities and estimate the nuclear structure function $F_2^A(x,Q^2)$, as well as  the longitudinal and charm contributions, using a generalization for nuclear targets of the Iancu-Itakura-Munier model which describes the $ep$ HERA data  quite well. Moreover, we investigate the behavior of the logarithmic slopes of the total and longitudinal structure functions in the kinematical region of the future electron - ion collider eRHIC.

\end{abstract}
\maketitle
\vspace{1cm}

In the high energy limit, perturbative Quantum Chromodynamics (pQCD) predicts that the 
small-$x$ gluons in a hadron wavefunction should form a Color Glass Condensate (CGC), which is 
 described by an infinite hierarchy of  coupled evolution equations for the correlators of 
Wilson lines \cite{VENUGOPALAN,BAL,CGC,WEIGERT}.  In the absence of correlations, the first 
equation in the Balitsky-JIMWLK hierarchy decouples and is then equivalent to the equation 
derived independently by Kovchegov within the dipole formalism \cite{KOVCHEGOV}. 
The Balitsky-Kovchegov (BK) equation describes the energy evolution of the dipole-target scattering amplitude ${\cal{N}}(x,\rr)$. Although a complete analytical solution is still lacking, its main properties are known (for recent 
reviews see, e.g. \cite{iancu_raju,anna_review,weigert_review,kov_jamal_review}): (a) for the interaction of a small dipole ($\rr
\ll 1/Q_{\mathrm{s}}$), ${\cal{N}}(\rr) \approx \rr^2$, implying  that
this system is weakly interacting; (b) for a large dipole
($\rr \gg 1/Q_{\mathrm{s}}$), the system is strongly absorbed and therefore 
${\cal{N}}(\rr) \approx 1$.  This property is associated  to the
large density of saturated gluons in the hadron wave function.  
Furthermore, several groups have studied the numerical solution of the BK equation \cite{solBK,weigert_ruma,albacete} 
and confirmed many of the theoretical predictions. In particular, the studies 
presented in  \cite{weigert_ruma,albacete} have 
demonstrated that the BK  solution for fixed constant coupling preserves the atomic number dependence of the saturation scale 
present in the initial condition, while for running $\alpha_s$ this dependence is reduced with increasing rapidity, as predicted by Mueller in Ref. \cite{mueller_a}.

 The search for signatures of parton saturation effects has been an
active subject of research in the last years (for recent reviews see, e.g. 
\cite{iancu_raju,kov_jamal_review,vicmag_mpla}).
On one hand, it has  been observed that the HERA data at small $x$ and low $Q^2$ can be
successfully described with the help of saturation models \cite{GBW,bgbk,kowtea,iancu_munier,fs}, with the  experimental results for the total cross section \cite{scaling} and   
inclusive charm production \cite{prl} presenting  the property of  geometric scaling.  On the other hand, the recently  observed \cite{BRAHMSdata}
suppression of high $p_T$ hadron yields at forward rapidities in dAu collisions at RHIC 
has the behavior  anticipated on the basis of CGC ideas \cite{cronin,jamal,kkt05,Dumitru}. All these results provide strong evidence for the CGC physics at HERA and RHIC. However, more definite conclusions are not possible due to the small value of the saturation scale in the kinematical range of HERA and due to the complexity present in the description of $d Au$ collisions, where we need to consider the substructure of projectile and target as well as the fragmentation of the produced partons. As a direct consequence, other models are able to describe the same set of data (See e.g. Refs. \cite{forshaw_twopomeron,hwa}). In order to discriminate between these different models and test the CGC physics, it would be very important to consider an alternative search. An ideal place are the future electron-nucleus colliders, which probably could determine  whether parton distributions saturate and constrain the behavior of the nuclear gluon distribution. This expectation can easily be understood if we assume the empirical parameterization 
$Q_s^2 = A^{\frac{1}{3}} \times Q_0^2 \, (\frac{x_0}{x})^{\lambda}$, with the parameters 
$Q_0^2 = 1.0$ GeV$^2$, $x_0 = 0.267 \times 10^{-4}$ and $\lambda = 0.253$ as in 
Ref. \cite{iancu_munier}. We can observe that, while in the proton case we need very 
small values of $x$ to obtain large values of $Q_s^2$, in the nuclear case a similar value 
can be obtained for values of $x$ approximately two orders of magnitude greater.
Consequently, nuclei are an efficient amplifier of  parton densities. The parton density that would be accessed 
in an electron - ion collider would be equivalent to that obtained in an electron - proton collider at energies 
that are at least one order of magnitude higher than at HERA \cite{raju_eA}.

Recently, an electron - ion collider at RHIC has been proposed in order to explore the relevant physics of  polarized and unpolarized electron-nucleus collisions \cite{raju_annual}. In particular, this collider will explore the high density regime of 
QCD even though its $x - Q^2$ range will be somewhat less extensive than  achieved at HERA. For instance, with energies $\sqrt{s} = 60 - 100$ GeV, one will access $x \approx 10^{-4} - 10^{-3}$ for $Q^2 \approx 1 - 10 $ GeV$^2$, respectively. However, the saturation scale will be approximately 4.0 GeV$^2$ at small $x$, low $Q^2$ and $A = 197$. Furthermore, as pointed in Ref. \cite{raju_annual}, in principle all the inclusive and semi-inclusive observables that were studied at HERA can be studied at eRHIC. This collider is expected to have statistics high  
enough to allow for the determination of  the logarithmic slopes with respect to $x$ and $Q^2$ of the total and longitudinal structure functions. In particular, the longitudinal structure 
function is expected to be measured for the first time in the kinematical regime of small 
$x$, since the electron - ion collider will be able to vary the energies of both the electron 
and ion beams. It  will be possible to check  predictions made by CGC inspired models  
(which have been extensively tested at HERA) for the behavior of these observables.

In this paper we study the behavior of  the total, longitudinal and charm structure functions
 in the kinematical region which will be probed in electron-ion collisions at RHIC 
considering a generalization for nuclear targets of the saturation model proposed by Iancu, 
Itakura and Munier (IIM model). Moreover, we estimate the logarithmic slopes of the total 
and longitudinal structure functions at different values of the atomic number. We hope that 
our results contribute to the planning of future $eA$ experiments (for previous studies see 
Refs. 
\cite{AYALAS,victor_slope1,victor_prc,victor_slope2,levin_mcl,victor_ayala,armesto_braun,levin_lub0,armesto,victor_mag_heavy1,armesto_epjc,bartels_photo,weigert_prl,levin_lub,victor_mag_heavy2,victor_mag_vector,armesto_prl}).

We start from the space-time picture of the electron-proton/nuclei
processes \cite{dipole}.
In the rest frame of the target, the QCD description of
DIS at small $x$ can be
interpreted as a two-step process. The virtual photon (emitted
by the incident electron) splits into a $q\bar{q}$ dipole which
subsequently interacts with the target.  In terms
of  virtual photon-target cross sections  $\sigma_{T,L}$
for the transversely and longitudinally  polarized photons, the $F_2$ structure function is 
given by  \cite{dipole}
\be
\label{eq:1}
F_2(x,Q^2)\,=\,\frac{Q^2}{4 \pi^2 \alphaem} \,(\sigma_T\,+\,\sigma_L)
\ee
and
\be
\label{eq:2}
\sigma_{T,L}\,=\,  \int d^2{\rr}\, dz\, |\Psi_{T,L}(\rr,z,Q^2)|^2\,\, \sigma_{dip}(x,\rr),
\ee
where $\Psi_{T,L}$ is the light-cone  wave function of the virtual photon
and $\sigma_{dip}$ is the  dipole cross section
describing  the interaction of the $q\bar{q}$  dipole with the target.  In
equation (\ref{eq:2})
$\rr$ is the transverse separation of the $q\bar{q}$ pair
and $z$ is the photon  momentum fraction carried by the quark (For details see  e.g. Ref. \cite{PREDAZZI}) .

\begin{figure}
\centerline{
{\psfig{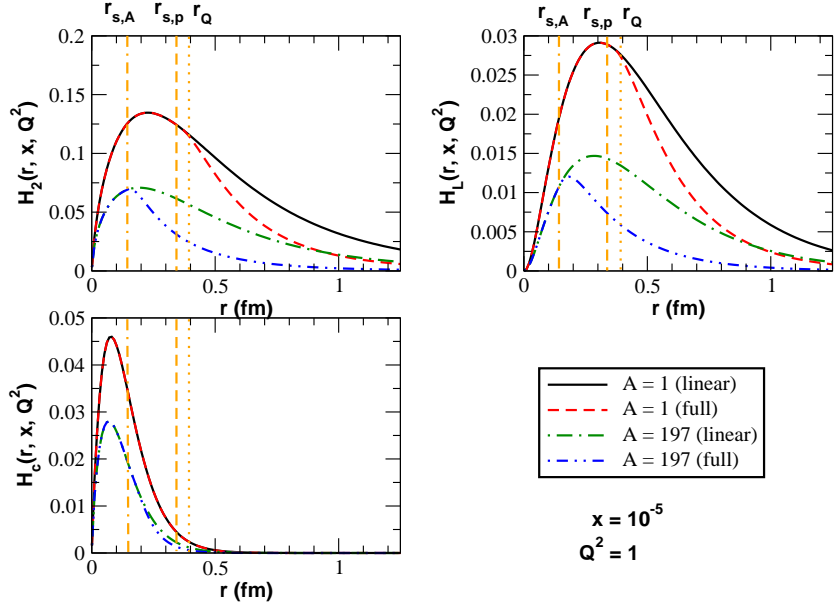}}}
\caption{The $\rr$-dependence of the photon-nucleus overlap functions, normalized by $A$, for different values of the atomic number ($x = 10^{-5}$ and $Q^2$ = 1 GeV$^2$).}
\label{fig1}
\end{figure}

The dipole hadron cross section $\sigma_{dip}$  contains all
information about the target and the strong interaction physics.
In the Color Glass Condensate  formalism \cite{BAL,CGC,WEIGERT}, 
$\sigma_{dip}$ can be
computed in the eikonal approximation and is given by:
\begin{eqnarray}
\sigma_{dip} (x,\rr)=2 \int d^2 \rb \, {\cal{N}}(x,\rr,\rb)\,\,,
\end{eqnarray}
where ${\cal{N}}$ is the  dipole-target forward scattering amplitude for a given impact 
parameter $\rb$  which encodes all the
information about the hadronic scattering, and thus about the
non-linear and quantum effects in the hadron wave function. The
function ${\cal{N}}$ can be obtained by solving the BK (JIMWLK) evolution
equation in the rapidity $Y\equiv \ln (1/x)$. 
It is useful to assume that the impact parameter dependence of $\cal{N}$ can be factorized as 
${\cal{N}}(x,\rr,\rb) = {\cal{N}}(x,\rr) S(\rb)$, so that 
$\sigma_{dip}(x,\rr) = {\sigma_0} \,{\cal{N}}(x,\rr)$, with $\sigma_0$ being   a free 
parameter  related to the non-perturbative QCD physics. 
Several models for the dipole cross section have been used in the literature in order to fit 
the HERA data \cite{GBW,bgbk,kowtea,iancu_munier,fs}. Here we will consider only the model proposed in Ref. \cite{iancu_munier} 
where the  dipole-target forward scattering amplitude  was parametrized 
as follows,
\begin{eqnarray}
{\cal{N}}(x,\rr) =  \left\{ \begin{array}{ll} 
{\mathcal N}_0\, \left(\frac{\rr\, Q_s}{2}\right)^{2\left(\gamma_s + 
\frac{\ln (2/\rr Q_s)}{\kappa \,\lambda \,Y}\right)}\,, & \mbox{for $\rr Q_s(x) \le 2$}\,,\\
 1 - \exp^{-a\,\ln^2\,(b\,\rr\, Q_s)}\,,  & \mbox{for $\rr Q_s(x)  > 2$}\,, 
\end{array} \right.
\label{CGCfit}
\end{eqnarray}
where the expression for $\rr Q_s(x)  > 2$  (saturation region)   has the correct 
functional
form, as obtained either by solving the BK equation 
\cite{BAL,KOVCHEGOV}, 
or from the theory of the Color Glass Condensate  \cite{iancu_raju}. Hereafter, 
we label the model above by IIM. The coefficients $a$ and $b$ are determined from the 
continuity conditions of the dipole cross section  at $\rr Q_s(x)=2$. The coefficients 
$\gamma_s= 0.63$ and $\kappa= 9.9$  are fixed from their LO BFKL values and $\sigma_0 = 2 \pi R_p^2$, where $R_p$ is the proton radius. In our 
further calculations we shall  use the parameters $R_p=0.641$ fm, $\lambda=0.253$, 
$x_0=0.267\times 10^{-4}$ and ${\mathcal N}_0=0.7$, which give the best fit result. 
Recently, this model has also been used in phenomenological studies of  vector meson 
production \cite{Forshaw1} and  diffractive processes \cite{Forshaw2} at HERA as well 
as for the description of the longitudinal structure function \cite{vicmag_fl}.

 { Some comments related to IIM model are in order here. Firstly, it is important to emphasize that this model is constructed by smoothly interpolating between two limiting cases which are analytically under control, but have leading order accuracy. 
The first line of Eq. (\ref{CGCfit}) is obtained from the solution of the BFKL equation via a saddle point approximation, valid for very high energies and very small dipole sizes. The so obtained solution is then further expanded under the assumption that the dipole sizes are close to the saturation radius.
Secondly, it is valid only in a  limited range of virtualities, such that the DGLAP evolution can be disregarded and the scaling solution of the BFKL equation can be used. Moreover, the free parameters in the IIM model have been determined without including the charm, {\it i.e.} considering only three active flavors. Consequently, our predictions for the nuclear charm structure function should be considered as a rough estimate of this observable. However, we believe that our main conclusions related to $F_2^{c,A}$ are not modified if a 
new fit including the charm is performed. }


\begin{figure}
\centerline{
{\psfig{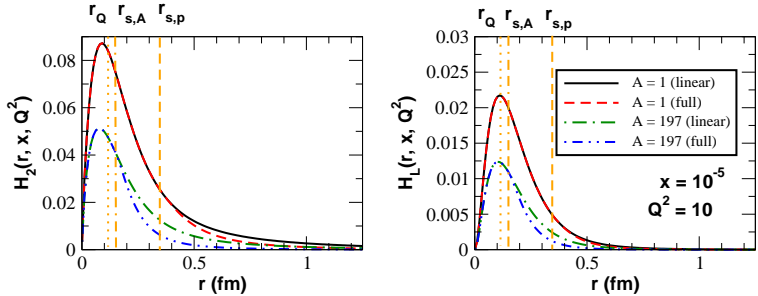}} }
\caption{The $\rr$-dependence of the photon-nucleus overlap functions, normalized by $A$, for different values of the atomic number ($x = 10^{-5}$ and $Q^2$ = 10 GeV$^2$).  }
\label{fig2}
\end{figure}

We generalize the IIM model for nuclear collisions  assuming  the following basic transformations: 
$\sigma_0 \rightarrow \sigma_0^A =  A^{\frac{2}{3}} \times \sigma_0$ and $Q_s^2(x) \rightarrow Q_{s,A}^2 =  A^{\frac{1}{3}} \times  Q_s^2(x)$. Moreover, in order to estimate the contribution of the saturation physics in what follows we present a comparison between the full IIM model and the predictions  from  linear physics, obtained extrapolating the expression of ${\cal{N}}(x,\rr)$, valid for  $\rr Q_s(x) \le 2$, 
for all kinematical range. { In our calculations  the impact parameter dependence of the scattering amplitude, which is mainly associated to non-perturbative physics, is disregarded. Basically, following Ref. \cite{iancu_munier}, we shall treat the nucleus as a homogeneous disk of radius $R_A$. Consequently, with this model we cannot discuss the expansion of the transverse size of the target with increasing energy. Furthermore, we are assuming that the $A^{\frac{1}{3}}$-dependence of the nuclear saturation scale is preserved by the evolution. This assumption is valid in the fixed coupling case, where the $A$ scaling of the initial condition  survive without change the evolution. On the other hand, for a running coupling the $A^{\frac{1}{3}}$ scaling holds only in a limited kinematical range, with the nuclear dependence becoming weaker at larger energies \cite{mueller_a}. In this case, we can  expect that the  $A^{\frac{1}{3}}$  dependence of the nuclear saturation scale will be slightly  modified at eRHIC.   }
It is important to emphasize that more sophisticated generalizations to the nuclear case can be used (See e.g. Refs. \cite{kowtea,armesto,bartels_photo,armesto_prl}). Consequently, this work should be considered as an exploratory study which has as main goal to present a semiquantitative estimate of the CGC effects in the future $eA$ collider. In a full 
calculation we must  use the solution of the BK equation, obtained without neglecting the impact parameter dependence as well as an initial condition constrained by current experimental lepton-nucleus data.

\begin{figure}
\centerline{
{\psfig{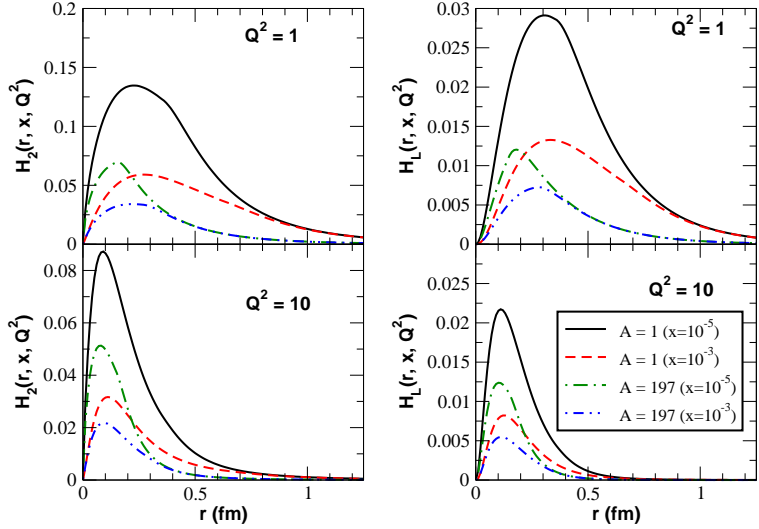}} }
\caption{The $\rr$-dependence of the photon-nucleus overlap functions, normalized by $A$, for different values of the atomic number, $x$ and $Q^2$.}
\label{fig3}
\end{figure}

Before presenting our results for the total, longitudinal and charm structure function, we can investigate the mean dipole size dominating each of these  cross sections. We define  the photon-nucleus overlap function, { normalized  by the atomic number $A$, as follows }
\begin{eqnarray}
H_i \,(\rr,x,Q^2)  = \frac{ 2\pi \rrn}{A} \, \int dz\, |\Psi_i
(z,\,\rr, m_f, Q^2)|^2 \, \sigma_{dip} (x,\rr,A)\,.
\label{overlap}
\end{eqnarray}
where $i = T, \,L$ characterizes  transverse and longitudinal photons. In particular, we also calculate the overlap function associated to the total structure function, $H_2 \,(\rr,x,Q^2) \equiv H_T \,(\rr,x,Q^2) + H_L \,(\rr,x,Q^2)$, and the overlap function associated to the charm structure function, $H_c \,(\rr,x,Q^2)$, which is calculated using $m_f = m_c = 1.5$ GeV. In Figs. \ref{fig1}, \ref{fig2} and \ref{fig3} we show the distinct overlap functions 
(normalized by $A$) as a function of the dipole size for different values of $x$, $A$ and 
$Q^2$.  The first aspect that should be emphasized is that  although the overlap functions 
have been normalized by $A$, they are strongly $A$ dependent. This behavior is expected when 
we consider the full prediction of the IIM model. However, our results demonstrate that this 
dependence is also present when we calculate the overlap functions using the linear approximation. It is associated to the $[Q_{s,A}^2(x)]^{\gamma_{eff}}$ dependence of the dipole scattering amplitude, where $\gamma_{eff} = \gamma_s + 
\frac{\ln (2/\rr Q_s)}{\kappa \,\lambda \,Y}$ in the IIM model. For $\gamma_{eff} = 1$ we will have an $A^{\frac{1}{3}}$-dependence for ${\cal{N}}$, which combined with the $A^{\frac{2}{3}}$-dependence of $\sigma_0$ implies a linear $A$ dependence for the dipole cross section in the linear regime. If normalized by $A$ we will obtain an $A$ independent overlap function in the linear regime. However, since $\gamma_{eff} < 1$ in the IIM model,  the overlap function (and the corresponding observables) has an $A^{\frac{\gamma_{eff} - 1}{3}}$ dependence, i.e. it decreases with increasing  atomic number. This behavior is observed in the figures.
In Fig. \ref{fig1} we estimate the overlap functions for fixed $x$ ($= 10^{-5}$) and $Q^2$ (= 1 GeV$^2$) and two values of the atomic number.
{ In order to illustrate our results we define the quantities $\rr_{s,A} \equiv 2/Q_{s,A}$, $\rr_{s,p} \equiv 2/Q_{s,p}$ and $\rr_Q \equiv 2/Q$, which are directly associated to the nuclear, proton saturation scale and photon virtuality, respectively. The value of these quantities is indicated in Figs. \ref{fig1} and \ref{fig2} by vertical lines. We have $\rr_{s,A} < \rr_{s,p} < \rr_Q$ at $Q^2 = 1$ GeV$^2$, while   $\rr_Q < \rr_{s,A} < \rr_{s,p}$ at $Q^2 = 10$ GeV$^2$. } 
 We can see that the charm overlap function is peaked at approximately $\rr \approx 0.07$ fm, which agrees with the theoretical expectation that the $c\overline{c}$ pair has a typical transverse size { $\approx 1/\mu$, where $\mu \equiv \sqrt{ Q^2 + 4 m_c^2}$} (See similar discussion in Ref. \cite{victor_mag_heavy1}). Therefore, the main contribution to the cross section comes from the small dipole sizes, i. e. from the region where the saturation effects are small (linear regime). This expectation is confirmed by  the behavior of the overlap function $H_c$ which is the same in  
the linear and full predictions for fixed $A$. Therefore, we should expect that the modifications in the charm structure function due to saturation effects will be small.
{ This is expected since the typical scale for charm production,  $\mu^2$, is larger than the saturation scale $Q_{s,A}^2$ in all kinematical range of eRHIC, {\it i.e.} at this collider it is expected that the linear regime dominates the  heavy quark  production (See discussion in Ref. \cite{prl})}.  On the other hand, for light quark production a broader $\rr$ distribution is obtained, peaked at large values of the pair separation, { $\rr \approx \rr_{s,p} \, (\rr_{s,A})$ at $A$ = 1 (197)},  implying that saturation effects contribute significantly in this case. 
This same feature is observed in the behavior of the overlap functions $H_2$ and $H_L$ presented in Fig. \ref{fig1}. In this case we can see 
that the saturation effects suppress the contribution of large dipole size, as expected theoretically. Moreover, these effects become more important for smaller values of $x$ and larger $A$. We can observe that the area under the curve is significantly reduced by the saturation effects, which implies that the associated observable will be strongly modified by these effects. In Fig. \ref{fig2} we present the behavior of $H_2$ and $H_L$ for $Q^2 = 10$ GeV$^2$.  In this case we observe that  the distributions peak at smaller values of the dipole size, {\bf $\rr \approx \rr_Q$}, with the contribution of large dipole size being reduced. It implies that  for $A = 1$ the full and linear predictions are identical. For large nuclei ($A = 197$)  saturation effects still 
contribute, which implies a reduction of the area under the curve and a modification of the associated observable. The charm overlap function (not shown) has identical behavior for the linear and full predictions and the two values of the atomic number. We can conclude that the saturation effects are strongly reduced for large values of $Q^2$.

\begin{figure}
\vspace*{0.5cm}
\centerline{\psfig{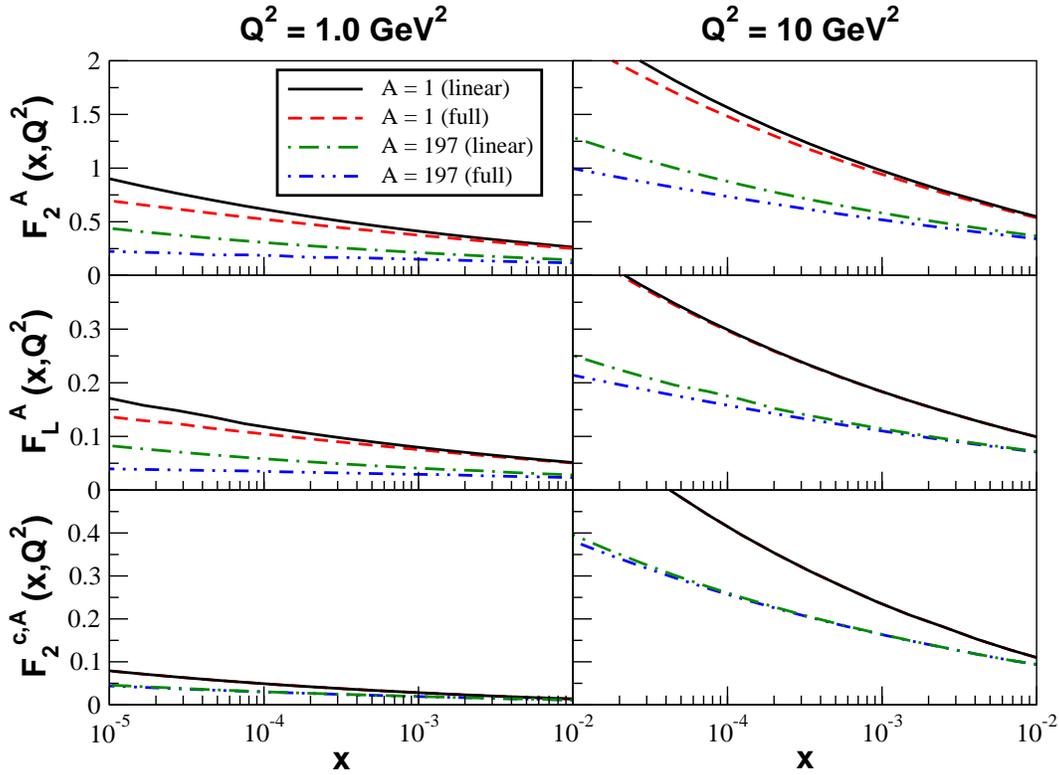}}
\caption{Nuclear structure functions as a function of $x$ for different values of $A$ and $Q^2$.}
\label{fig4}
\end{figure}

\begin{figure}
\centerline{
\begin{tabular}{ccc}
{\psfig{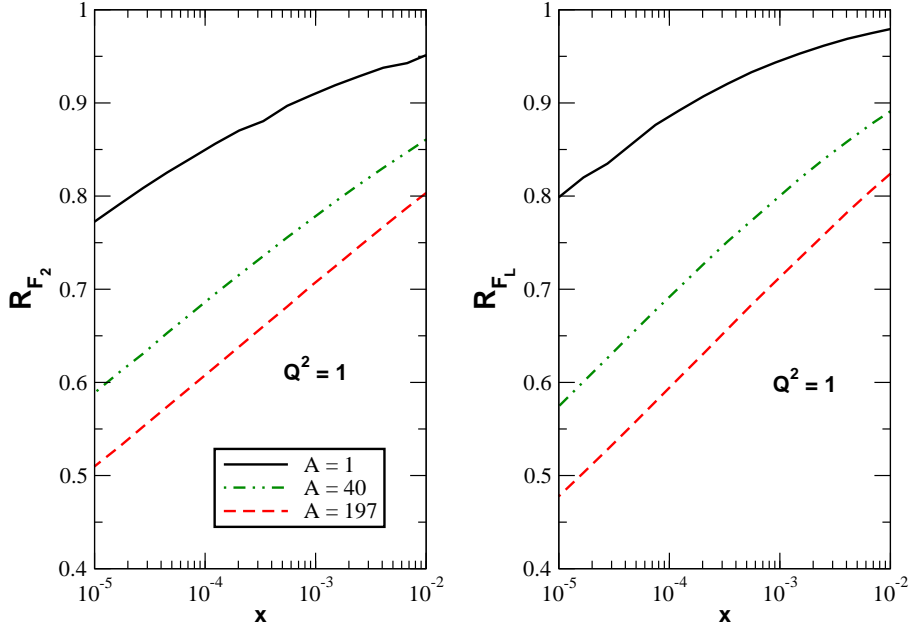}} & \,\,\,\,\,\, &  
\end{tabular}}
\caption{Ratio between the full and linear predictions for the different nuclear structure functions.}
\label{fig5}
\end{figure}

\begin{figure}
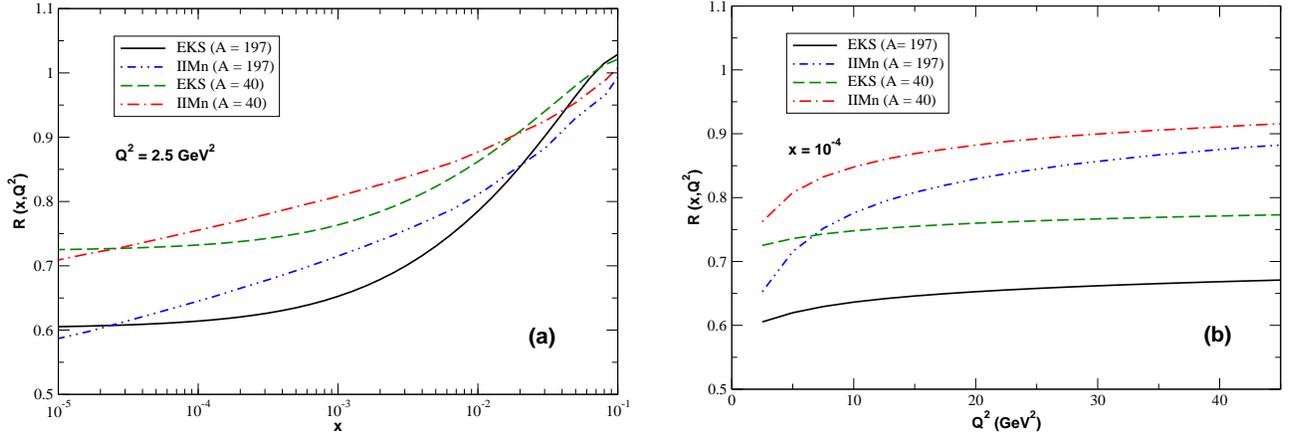

\vspace*{0.5cm}
\centerline{
\begin{tabular}{ccc}
{\psfig{figure=raznucx.eps,width=8.3cm}} & \,\,\,\,\,\, & {\psfig{figure=raznucq.eps,width=8.cm}} 
\end{tabular}}
\caption{Ratio $R (x,Q^2) \equiv F_2^{A}(x,Q^2)/F_2^{p}(x,Q^2)$ as a function of (a) $x$ and (b) $Q^2$. The predictions of the EKS parameterization are shown for comparison.  }
\label{fig5b}
\end{figure}

In Fig. \ref{fig3} we present the $H_2$ and $H_L$ overlap functions for two values of $x$ and $Q^2$. In this figure we show only the full predictions. As discussed before: (a) increasing $Q^2$ the distributions peak at small values of the dipole size; and (b) the area under the curve is reduced increasing the atomic number. The main aspect of this figure is that it allows to analyse the $x$-dependence of the overlap function. We observe that decreasing $x$ the overlap function grows, with the growth  being smaller for large nuclei. Consequently, we expect that the associated observables increase at small $x$, with a smaller slope for $A = 197$. These expectations are confirmed in Fig. \ref{fig4}, where we present the $x$-dependence of the total, longitudinal and charm structure functions. We can see that the full and linear predictions for the charm structure function are identical, having the $A$-dependence characteristic of the IIM model. On the other hand, the behavior of $F_2^A$ and $F_L^A$ is strongly modified by the saturation physics, with the effect decreasing for larger $Q^2$.  In order to obtain a more precise estimate of the modification  in the observables, in Fig. \ref{fig5} we present the ratios $R_{F_2}$ and $R_{F_L}$ between the full and linear predictions for $F_2$ and $F_L$, respectively. We consider three typical values of the atomic number. As expected, the contribution of the saturation physics increases at large nuclei and smaller values of $x$. In particular, for values of $x$ around $10^{-5}$ we predict a reduction of about 50$\%$ in the total and longitudinal structure functions. { In Fig. \ref{fig5b} we present the behavior of the ratio between the  nuclear and proton  structure functions, $R (x,Q^2) \equiv F_2^{A}(x,Q^2)/F_2^{p}(x,Q^2)$, as a function of $x$ and $Q^2$. For comparison, the predictions of the EKS parameterization \cite{eks}, which is a global fit of the nuclear experimental data using the DGLAP equation, is also presented.   In particular, in Fig. \ref{fig5b} (a) we show the behavior of this ratio as a function of $x$ at $Q^2 = 2.5$ GeV$^2$, assuming two different values of
 $A$: $197$ and $40$. We can see 
that, similarly to EKS parameterization, the ratio obtained using the IIM model generalized for nuclear targets (IIMn) decreases when $A$ increases. The main difference between the predictions is the behavior of the ratio $R$ at small $x$. While the EKS parameterization predicts that the ratio is constant in this limit, the IIMn one predicts that the ratio still  decreases at smaller values of $x$.  In Fig. \ref{fig5b} (b) we present the $Q^2$ dependence of the ratio at $x = 10^{-4}$. We find that while this behavior in the EKS parameterization is directly associated to the DGLAP evolution, in the IIMn prediction it is associated to the saturation  and geometric scaling regime. We see that the predictions differ significantly at small values of $Q^2$, where the saturation physics dominates.}

\begin{figure}
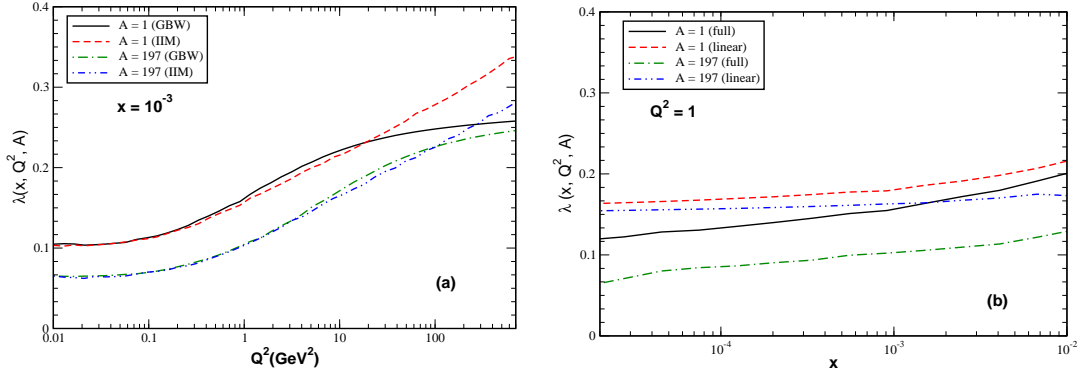

\centerline{
\begin{tabular}{ccc}
{\psfig{figure=lamx_3.eps,width=6.8cm}} & \,\,\, {\psfig{figure=lamq2_1.eps,width=7.cm}} &  \\
 &  & \\
\end{tabular}}
\caption{Effective intercept as a function of: (a)  $Q^2$ and (b) $x$.}
\label{fig6}
\end{figure}

Other observables of interest to study the CGC physics are the logarithmic slopes of $F_2$ and $F_L$ with respect to $x$ and $Q^2$. It is mainly motivated by the strict relation between the gluon distribution and the scaling violations of the total structure function at leading order in the DGLAP formalism \cite{derivada}. In the dipole formalism the scaling violations are directly related to the dipole cross section (See, e.g. Ref. \cite{ayala_froi})
\begin{eqnarray}
\frac{dF_2 (x,Q^2)}{d\log Q^2} \approx Q^2 \times \sigma_{dip}(x, \rr^2 = \frac{4}{Q^2}) \,\,.
\label{slopeq2}
\end{eqnarray}
Therefore, this observable can be useful to address the boundary between the linear and saturation  regimes \cite{victor_slope2}. This expectation is easily understood. As this observable is strongly dependent on the dipole cross section and it presents distinct behavior for $Q^2 > Q_s^2$ and $Q^2 < Q_s^2$, its experimental analysis would allow to test the $x$ and $A$ dependence of the saturation scale. On the other hand, the logarithmic derivative of $F_2$ with respect to $x$ is directly related with the power of growth of this structure function at small $x$.  Basically, if we parameterize the total structure function using $F_2^A (x,Q^2) = x^{- \lambda(x,Q^2,A)}$ we obtain that
\begin{eqnarray}
\frac{d\log F_2^A (x,Q^2)}{d\log 1/x} = \lambda(x,Q^2,A) \,\,,
\label{slopex}
\end{eqnarray}
i.e. this logarithmic slope is directly related to the effective Pomeron intercept (for a similar analysis in $ep$ collisions see, e.g. Ref. \cite{magno_kontros}). In Fig. \ref{fig6}(a) we present the $Q^2$ dependence of the effective intercept for $x = 10^{-3}$ and different values of the atomic number. For comparison, the prediction of the GBW model generalized for nuclear targets is also presented. For small values of $Q^2$ both models predict  a similar dependence for $\lambda$. The main difference occurs in the region of large values of $Q^2$ where  
their  predictions are  not expected to be valid. However, an $A$-dependence for $\lambda$ is observed,  $\lambda$ being smaller for large nuclei.
{ This behavior can be understood considering the  prediction of the IIM model for $F_2^A$ in the region  $\rr Q_s (x) \le 2$ (linear regime). In this case we obtain that the effective power $\lambda(x,Q^2,A)$ is proportional to  $\frac{\gamma_{eff} - 1}{3} \ln A / \ln (1/x)$ (plus positive $A$ independent terms), which is negative, since $\gamma_{eff} < 1$, and grows in modulus with $A$.} Moreover, this agrees with our previous results, 
where we have found that the growth of the nuclear structure function at small $x$ decreases at larger $A$.
 In \ref{fig6}(b) the $x$ dependence of the intercept for $Q^2 = 1$ GeV$^2$ is shown. In this case we only consider the IIM model and show its linear and full predictions. We can see that the linear predictions for $\lambda$  are similar. On the other hand, the saturation effects imply 
that $\lambda$  decreases  at small $x$, its reduction being stronger for $A = 197$.

\begin{figure}
\centerline{
\begin{tabular}{ccc}
{\psfig{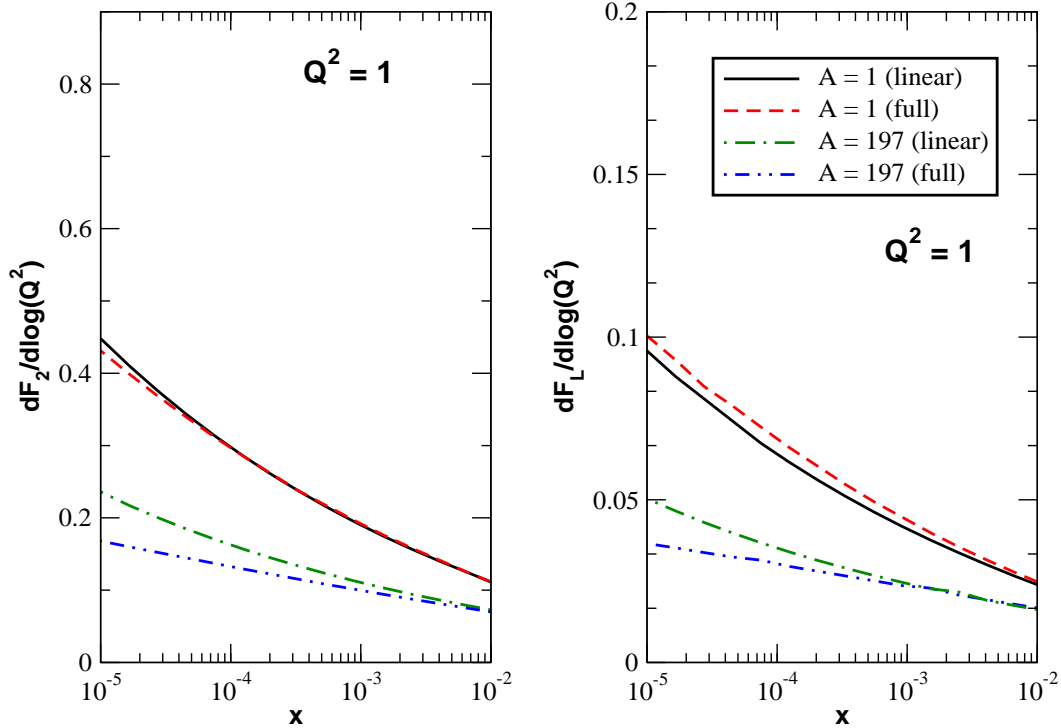}} & \,\,\,\,\,\, &  \\
& & \\
\end{tabular}}
\caption{Logarithmic slope with respect to $Q^2$ of the total and longitudinal structure functions.}
\label{fig7}
\end{figure}

\begin{figure}[h]
\centerline{
\begin{tabular}{ccc}
{\psfig{figure=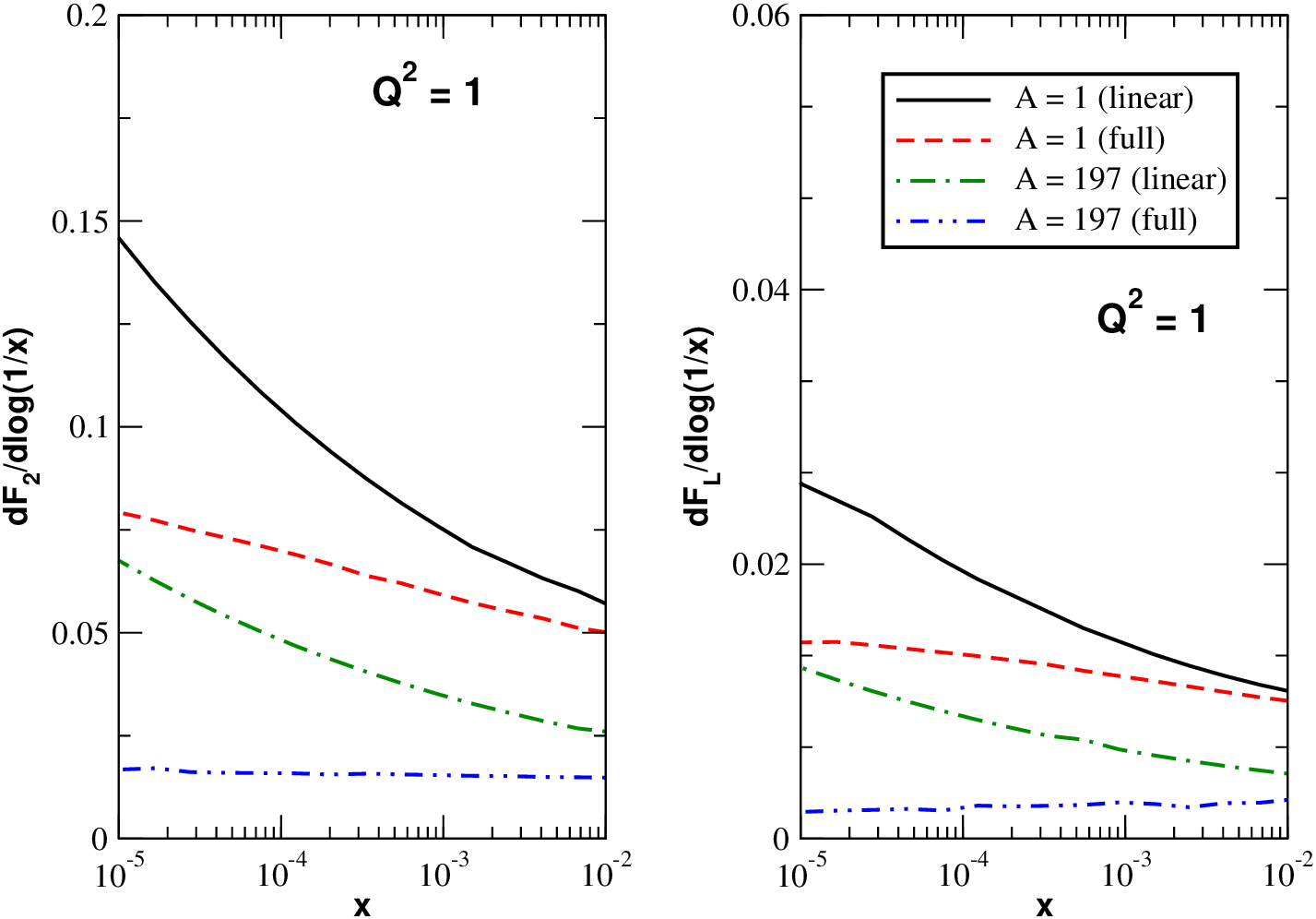,width=14.cm}} & \,\,\,\,\,\, &  \\
& & \\
\end{tabular}}
\caption{Logarithmic slope with respect to $x$ of the total and longitudinal structure functions.}
\label{fig8}
\end{figure}

In Fig. \ref{fig7} we present the $x$ dependence of the logarithmic derivatives of $F_2$ and $F_L$ with respect to $Q^2$. These derivatives, as well as $\lambda$, have been evaluated numerically using the DFRIDR routine \cite{nr}, which is based on the Richardson's deferred approach to the limit. We can see that both derivatives  have similar behavior, with the linear and full predictions being identical for large $x$. For the $F_2$ slope and $A = 1$, the difference between the linear and full predictions starts at $x \approx 10^{-4}$, increasing at smaller values of $x$. On the other hand, at $A = 197$ both predictions differ at values of $x$ smaller than  $\approx 10^{-2}$, with a large difference between the predictions at small $x$. For the $F_L$ slope and $A = 197$ we have a similar behavior, but with the difference between the predictions starting at   $x \approx 10^{-3}$. At $A = 1$ we can see that the full prediction is larger than the linear one for the $x$ range of the figure. We have found that at smaller values of $x$ the linear prediction becomes larger than the full one. Finally, in Fig. \ref{fig8} we present the $x$ dependence of the 
logarithmic derivatives of $F_2$ and $F_L$ with respect to $x$. We can see that while the linear predictions grow at small $x$, the full predictions present a smaller slope. In particular, at $A = 197$ these observables are almost $x$-independent when we consider the saturation effects.

As a summary, in this paper  
 we have studied the predictions of CGC physics for  electron - ion collisions at high energies, using a generalization for nuclear targets of the Iancu-Itakura-Munier model which describes the $ep$ HERA  quite well. We have  estimated the nuclear structure function $F_2^A(x,Q^2)$, as well as  the longitudinal and charm contributions. Moreover,   we have  investigated the behavior of the logarithmic slopes of the total and longitudinal structure functions in the kinematical region of the future electron - ion collider eRHIC.  Our results indicate that the experimental analysis of these observables in the future electron - ion collider could discriminate between linear and saturation physics, as well as constrain the behavior of the saturation scale.
Our analysis was restricted to inclusive observables. However, the CGC physics also  strongly modifies the behavior of exclusive observables, as verified, for instance, in diffractive processes 
at  HERA. Studies of  diffractive interactions in $eA$ interactions are still scarce. Some 
examples are those performed in  Refs. \cite{victor_mag_heavy1,victor_mag_vector}, where the diffractive photoproduction of heavy quark and vector mesons in $eA$ collisions were studied. In a forthcoming publication \cite{kuge} we calculate the nuclear diffractive structure function,
finding that the analysis of this observable can be useful to constrain the CGC physics.

\begin{acknowledgments}
  This work was  partially 
financed by the Brazilian funding
agencies CNPq, FAPESP and FAPERGS.
\end{acknowledgments}

\hspace{1.0cm}

\end{document}